\documentclass[a4paper,11pt]{article}
\usepackage{pos}

\title{Machine Can Automatically Discover Parametric Functions to Model HEP Data}

\author*[a]{Ho Fung Tsoi}
\author[a]{Dylan Rankin}
\author[b]{Cecile Caillol}
\author[c]{Miles Cranmer}
\author[d]{Sridhara Dasu}
\author[e]{Javier Duarte}
\author[f,g]{Philip Harris}
\author[a]{Elliot Lipeles}

\affiliation[a]{University of Pennsylvania, USA}
\affiliation[b]{European Organization for Nuclear Research (CERN), Switzerland}
\affiliation[c]{University of Cambridge, UK}
\affiliation[d]{University of Wisconsin-Madison, USA}
\affiliation[e]{University of California San Diego, USA}
\affiliation[f]{Massachusetts Institute of Technology, USA}
\affiliation[g]{Institute for Artificial Intelligence and Fundamental Interactions, USA}

\emailAdd{ho.fung.tsoi@cern.ch}

\abstract{In HEP data analyses, finding an adequate function to model binned data has largely relied on a manual process: guess a functional form by intuition, fit, examine, then repeat until successful. We show that this iterative process can be automated by a machine using symbolic regression, which performs a data-driven search over function space without requiring prior knowledge of what an adequate function should look like. We present the SymbolFit package, which pairs symbolic regression with uncertainty modeling to target HEP analysis use cases, and demonstrate it on the CMS and ATLAS Run 2 dijet spectra: 560 independent seeded runs across seven simple fit configurations generated over 1000 functions fitting the spectra with $\chi^2/\text{NDF}\approx 1$, and 111 of the runs rediscovered the very dijet and UA2 functions used in published dijet searches.}

\FullConference{
43rd International Conference on High Energy Physics (ICHEP 2026)\\
July 30 -- August 5, 2026\\
Natal, Brazil
}

\begin{document}
\maketitle

\section{Introduction}

Modeling binned data with smooth functions is ubiquitous in high-energy physics (HEP) data analysis, including deriving smooth scale factors to correct data-simulation mismodeling and modeling signal and background processes for hypothesis testing in new physics searches.
The main challenge is that most practical binned distributions do not have a ``true'' underlying function and hence an adequate functional form cannot be derived from first principles but only empirically.

The standard practice has remained the same for decades, even as of today.
An analyzer first guesses a functional form by intuition and writes down the parameterization by hand, then fits it to data and examines the residuals.
The first trials often fail with bad fit quality as the intuitive functional form is not adequate to describe a nontrivial distribution shape, and one needs to modify the form by, say, adding more parameters here and swapping the math operators there.
This repetitive process of form fine-tuning and refitting is empirical and creates a laborious loop that would sometimes impede the analysis progress.
The final candidate functions surviving this loop are often confined to a small corner of function space and highly fine-tuned to the analysis at hand.
They may no longer work for a future analysis after a luminosity increase or a trigger change, and a new family of functions has to be invented all over again.
Well-known examples are the ``dijet function'' used to model the QCD background in CMS and ATLAS Run 2 high-mass dijet resonance searches~\cite{CMS:2019gwf,ATLAS:2019fgd}, and the older ``UA2 function'' used alongside the dijet function in an ATLAS Run 2 low-mass dijet search~\cite{ATLAS:2018qto}:
\begin{equation}
    \label{eq:dijet}
    f_{\text{dijet}}(x) = \frac{p_0 (1-x)^{p_1}}{x^{p_2 + p_3 \ln x}}, \qquad f_{\text{UA2}}(x) = \frac{p_0\, e^{-p_2 x - p_3 x^2}}{x^{p_1}}, \qquad x=m_{jj}/\sqrt{s}.
\end{equation}

We propose symbolic regression (SR) to solve this problem in HEP.
SR is a machine learning method that performs a data-driven search over function space, where the functional form can be viewed as a ``trainable parameter'' that is dynamically changing during a fit to minimize the residual loss.
The function space is defined by a set of elementary math operators and the rules that connect them to construct a candidate function.
This eliminates the need to know what an adequate functional form looks like before fitting, as the search itself is automated.
We have previously introduced SymbolFit\footnote{{\url{https://github.com/hftsoi/symbolfit}}}~\cite{Tsoi:2024pbn}, an API that interfaces PySR~\cite{cranmerInterpretableMachineLearning2023} (SR search engine) and LMFIT~\cite{newville_2025_16175987} (nonlinear optimization and uncertainty estimation) for automating parametric modeling designed to target these HEP analysis applications.

In this contribution we use SymbolFit to run an experiment: can the data-driven search rediscover Eq.~\ref{eq:dijet} from the Run 2 dijet spectra published by CMS and ATLAS?
We show that not only does it succeed with a nontrivial rate, but it also readily discovers a wide range of different functions that model the spectra as well as the dijet and UA2 functions.

\section{SymbolFit}

SymbolFit streamlines three steps into a single pipeline for automation:
\begin{enumerate}
    \item \textbf{Function search.} PySR searches for adequate functions over a function space based on a multi-population evolutionary algorithm, without requiring a predefined functional form.
    
    \item \textbf{Re-optimization and uncertainty estimation.} Functions returned by PySR (and SR algorithms in general) are exact, carrying no uncertainty measure, which is an essential input to statistical analyses in HEP. Hence, the returned functions are parameterized and LMFIT is interfaced to re-optimize the numerical constants in each function and to compute uncertainties via the covariance matrix.
    
    \item \textbf{Evaluation.} All candidate functions are scored (e.g., $\chi^2/\text{NDF}$, $p$-value, RMSE, $R^2$), plotted with per-parameter uncertainty variation and total uncertainty coverage via parameter sampling. All results and plots are saved to easy-to-read output files for inspection.
\end{enumerate}

A search typically returns tens of functions, stochastically drawn from the large function space, and can be repeated with a different random seed to explore different batches of functions.

\section{Rediscovery setup}

We use two public datasets available on HEPData: the Run 2 observed dijet mass spectra at $\sqrt{s}=13$ TeV published by CMS~\cite{hepdata.91059} and ATLAS~\cite{hepdata.91126}, specifically Figure 5 from Ref.~\cite{CMS:2019gwf} and Figure 3a from Ref.~\cite{ATLAS:2019fgd}, respectively, where the dijet function in Eq.~\ref{eq:dijet} was used to model the spectra in the published results.
The task is to test whether, given only the data spectra, a data-driven search with SR can arrive at the same functional form that was empirically derived before.

\begin{table}[t]
  \centering
  \small
  \begin{tabular}{lll} \hline
    \textbf{Config variant} & \textbf{Function space}                                   & \textbf{Type} \\ \hline
    \texttt{ops-broad}      & operators $\{+,-,\times,/,\wedge,\exp,\ln,\tanh\}$                    & free-form \\
    \texttt{ops-log}        & operators $\{+,-,\times,/,\wedge,\ln\}$                               & free-form \\
    \texttt{ops-log-tight}  & same as \texttt{ops-log}, with $\wedge$-subtree sizes limited         & free-form \\
    \texttt{ops-minimal}    & operators $\{-,\times,/,\wedge\}$                                     & free-form \\
    
    \texttt{tpl-exp-x}      & $f = \exp(h(x))$, with $h$ built from $\{+,-,\times,\ln\}$            & template \\
    \texttt{tpl-pow-log}    & $f = g(x)^{\,h(\ln x)}$, with $g$ and $h$ built from $\{+,-,\times\}$ & template \\
    \texttt{tpl-exp-log}    & $f = \exp(h(\ln x, \ln(1-x)))$, with $h$ built from $\{+,-,\times\}$  & template \\ \hline
    
    \multicolumn{3}{l}{\textbf{Common PySR settings:} \texttt{niterations} $=200$ (free-form) / $100$ (template), \texttt{maxsize} $=40$,} \\ 
    \multicolumn{3}{l}{\texttt{model\_selection} $=$ \texttt{accuracy}, loss $=(y - \hat{y})^2 / \sigma^2$ per bin} \\ \hline
    \multicolumn{3}{l}{\textbf{Common SymbolFit settings:} $x=m_{jj}/\sqrt{s}$ fed directly, \texttt{input\_rescale\,$=$\,False},} \\
    \multicolumn{3}{l}{\texttt{scale\_y\_by\,$=$\,None}, \texttt{fit\_y\_unc\,$=$\,True}, \texttt{max\_stderr} $=20$} \\ \hline
  \end{tabular}
  \caption{The seven search configurations. Multiple nesting in operators is generally suppressed. Each config variant is used to perform 40 independent runs with different random seeds on each of the CMS and ATLAS dijet spectra (totaling $7\times 40\times 2=560$ seeded runs).}
  \label{tab:variants}
\end{table}

We define seven simple search configurations that set the function spaces, ranging from less constrained to more constrained space, summarized in Tab.~\ref{tab:variants}.
The four free-form configurations allow the operators to be combined freely when building candidate functions, and the three template configurations impose a functional template via PySR's \texttt{TemplateExpressionSpec} and search for the sub-expressions.
In general, nesting of operators is suppressed to avoid over-fitting.
Each configuration is independently run 40 times with different random seeds, identically on both datasets, giving a total of 560 seeded runs.
A run counts as a rediscovery if it returns at least one candidate that is algebraically equivalent to a member of the dijet or UA2 function families in Eq.~\ref{eq:dijet} and fits the spectrum with $\chi^2/\text{NDF}<2$.

\section{Results}

\begin{figure}[t]
  \centering
  \includegraphics[width=0.45\textwidth]{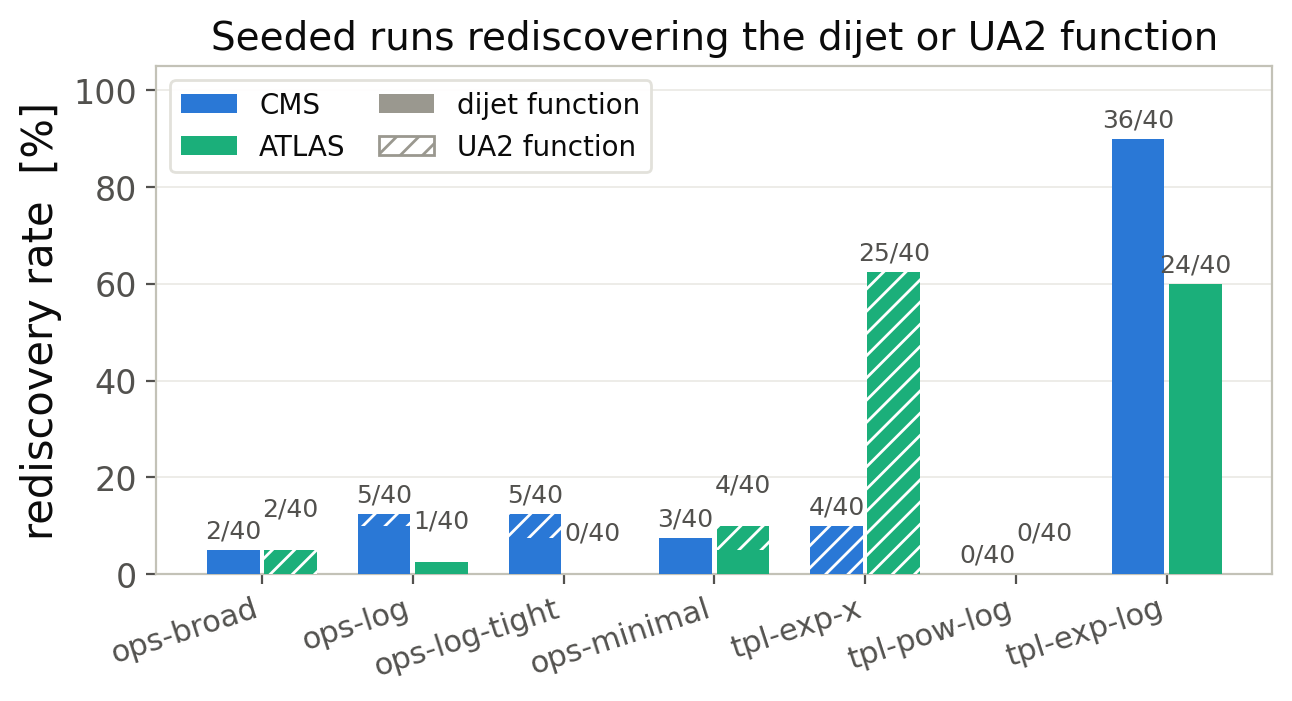}
  \caption{Fraction of seeded runs that rediscover the dijet function (solid) or the UA2 function (hatched). No run finds both, so in total 111 of the 560 runs rediscover one of the two. There are 40 runs per config variant per dataset.}
  \label{fig:hitrates}
\end{figure}

Fig.~\ref{fig:hitrates} shows the rediscovery rates.
Even nearly assumption-free searches find the dijet and UA2 functions.
For example, the free-form \texttt{ops-minimal} allowing $\{-, \times, /, \wedge\}$ has 3 of 40 runs and 2 of 40 runs rediscovering the dijet function from the CMS and ATLAS spectra, respectively.
When a template is imposed to guide the search by narrowing down the building blocks toward those of the dijet function, say \texttt{tpl-exp-log} with $f=\exp(h(\ln x, \ln(1-x)))$ searching for sub-expression $h$ built from $\{+,-,\times\}$, the rediscovery rates jump sharply to 36 and 24 out of 40 runs on the CMS and ATLAS spectra, respectively.
\texttt{tpl-exp-x} with $f=\exp(h(x))$ never finds the dijet function but finds the UA2 function in 4 of 40 and 25 of 40 runs on the CMS and ATLAS spectra, respectively.

\begin{figure}[t]
  \centering
  \includegraphics[width=0.55\textwidth]{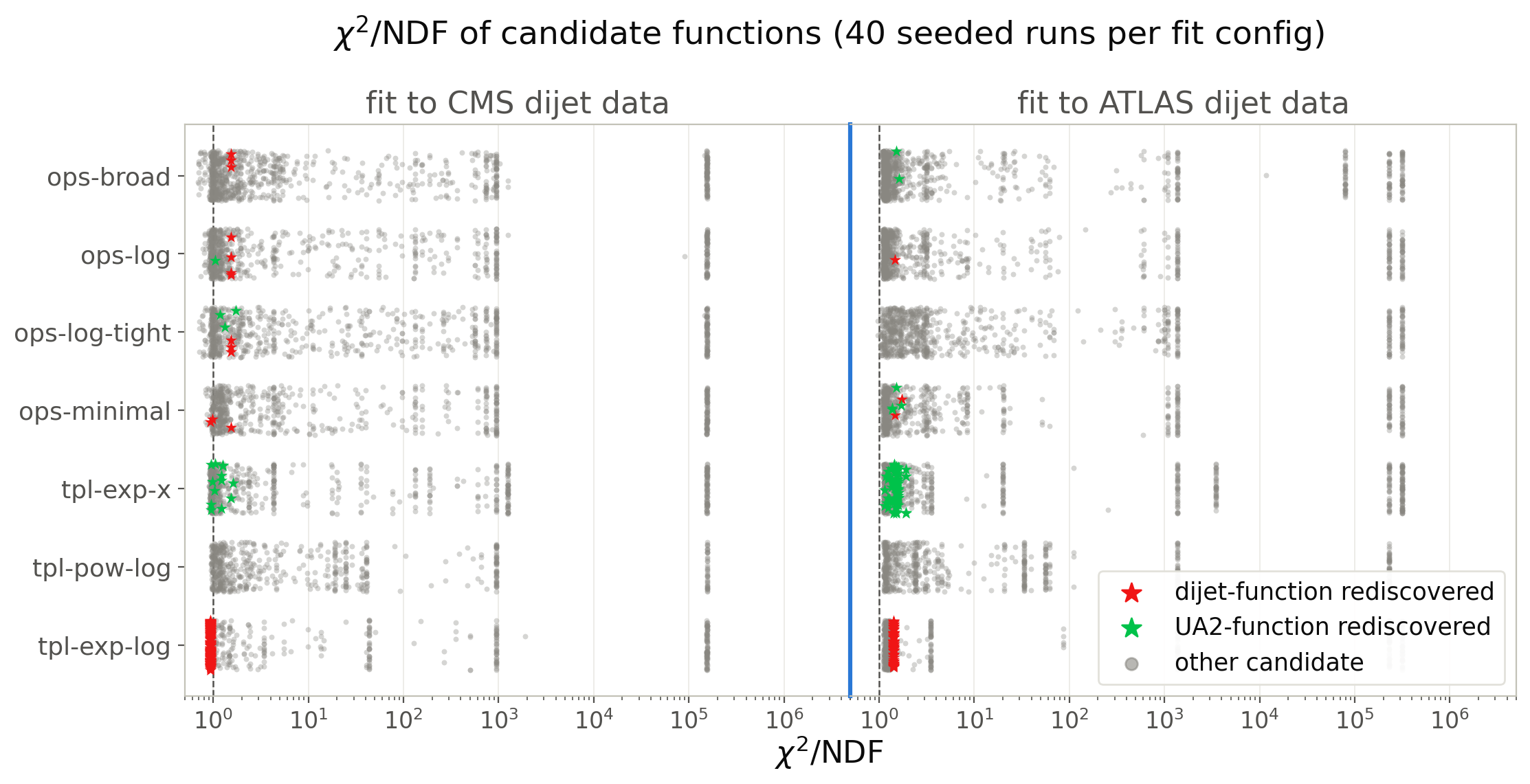}
  \caption{$\chi^2/\text{NDF}$ of all candidate functions returned by the
  $7\times 40\times 2=560$ seeded runs (each run produces tens of functions, and those with the same vertical position belong to the same run). Each marker represents a candidate function, and the red and green stars mark candidates equivalent to the dijet function and the UA2 function, respectively.}
  \label{fig:chi2}
\end{figure}

\begin{figure}[t]
  \centering
  \includegraphics[width=0.38\textwidth]{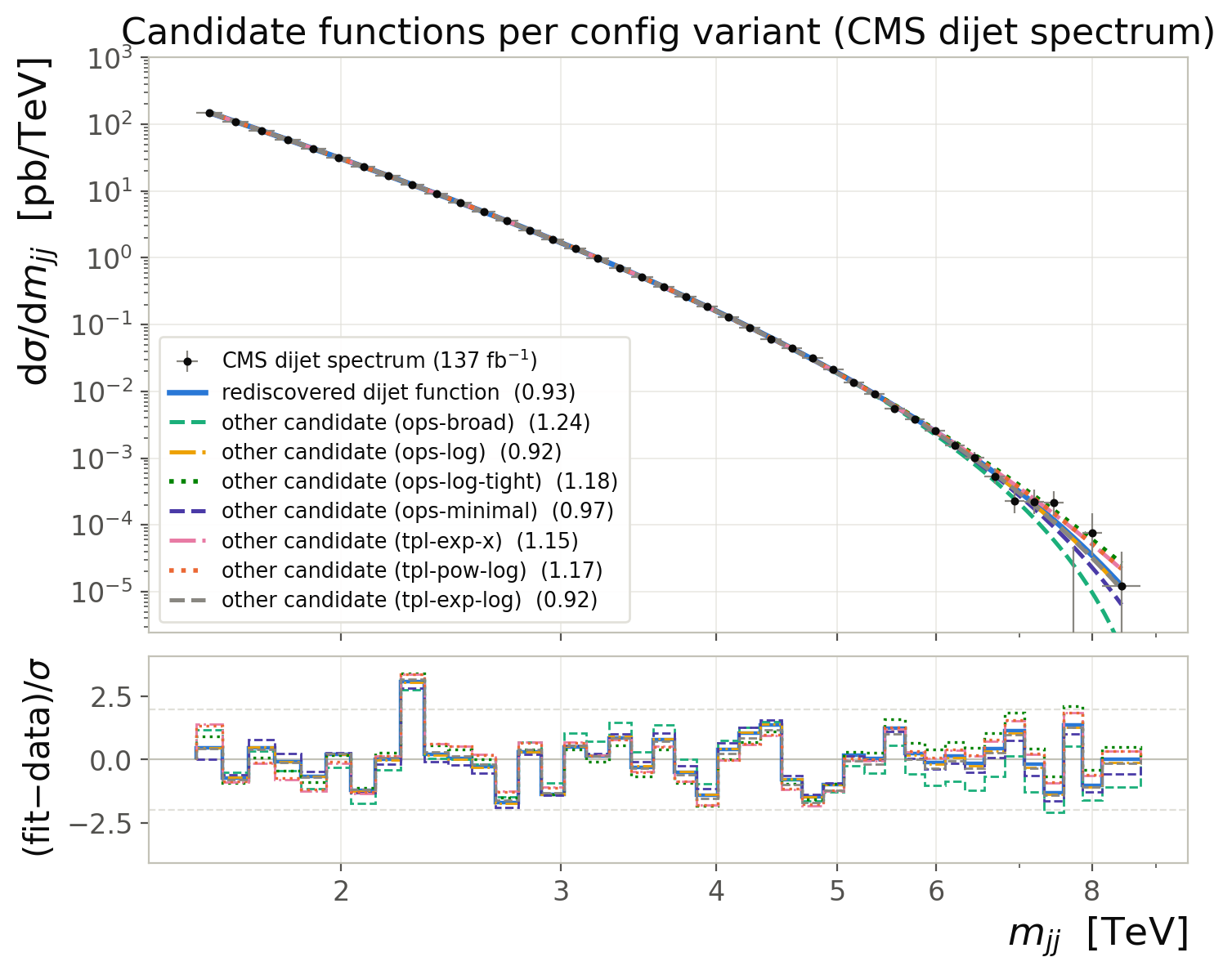}
  \includegraphics[width=0.38\textwidth]{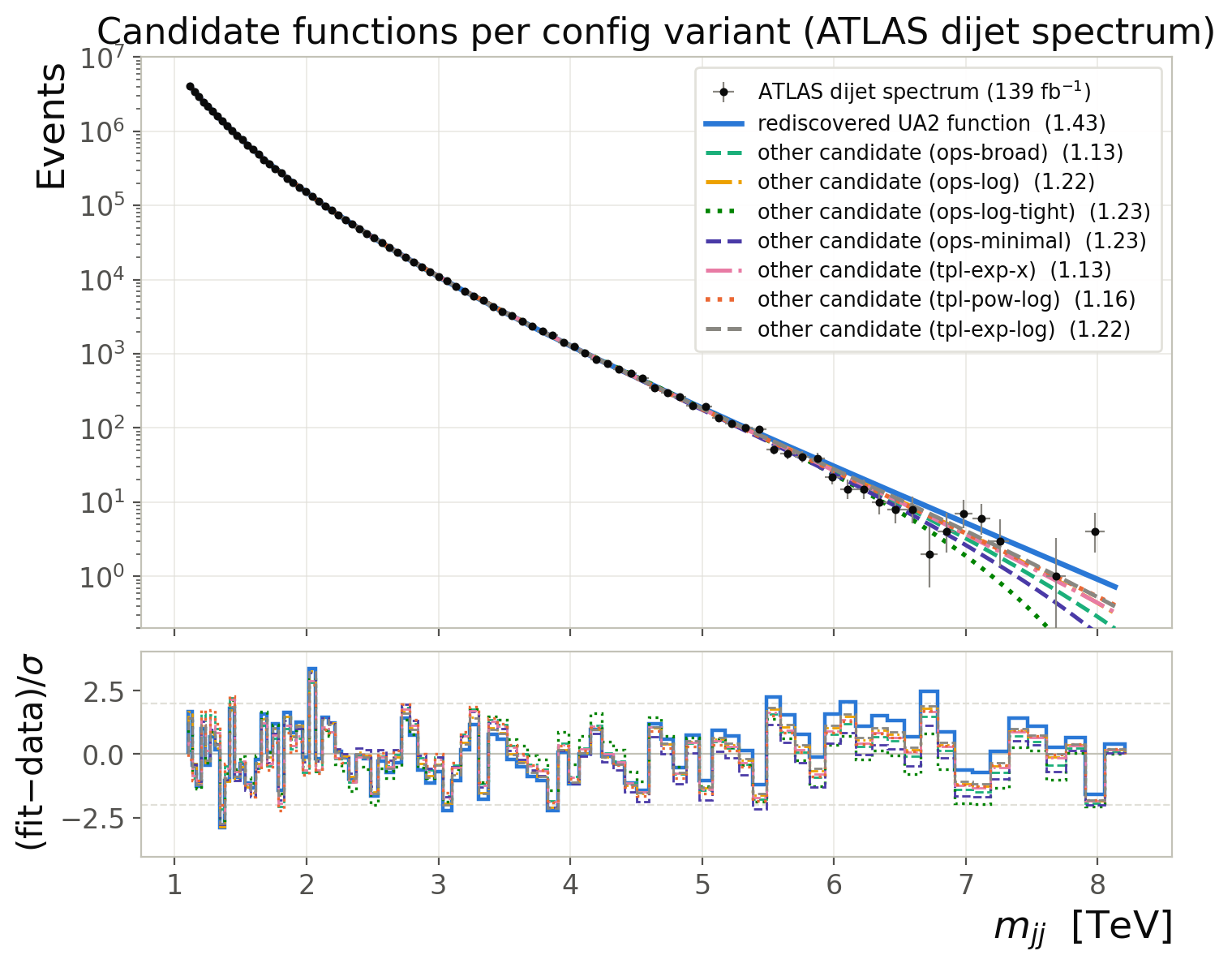}
  \caption{The rediscovered dijet function (CMS) and UA2 function (ATLAS), and an example candidate function (distinct from Eq.~\ref{eq:dijet}) from each search configuration, compared to the dijet spectra. Numbers in parentheses are $\chi^2/\text{NDF}$. Lower panels show the residual weighted by data uncertainty. The same candidates are listed in Tab.~\ref{tab:candidates}.}
  \label{fig:spectra}
\end{figure}

\begin{figure}[t]
  \centering
  \includegraphics[width=0.38\textwidth]{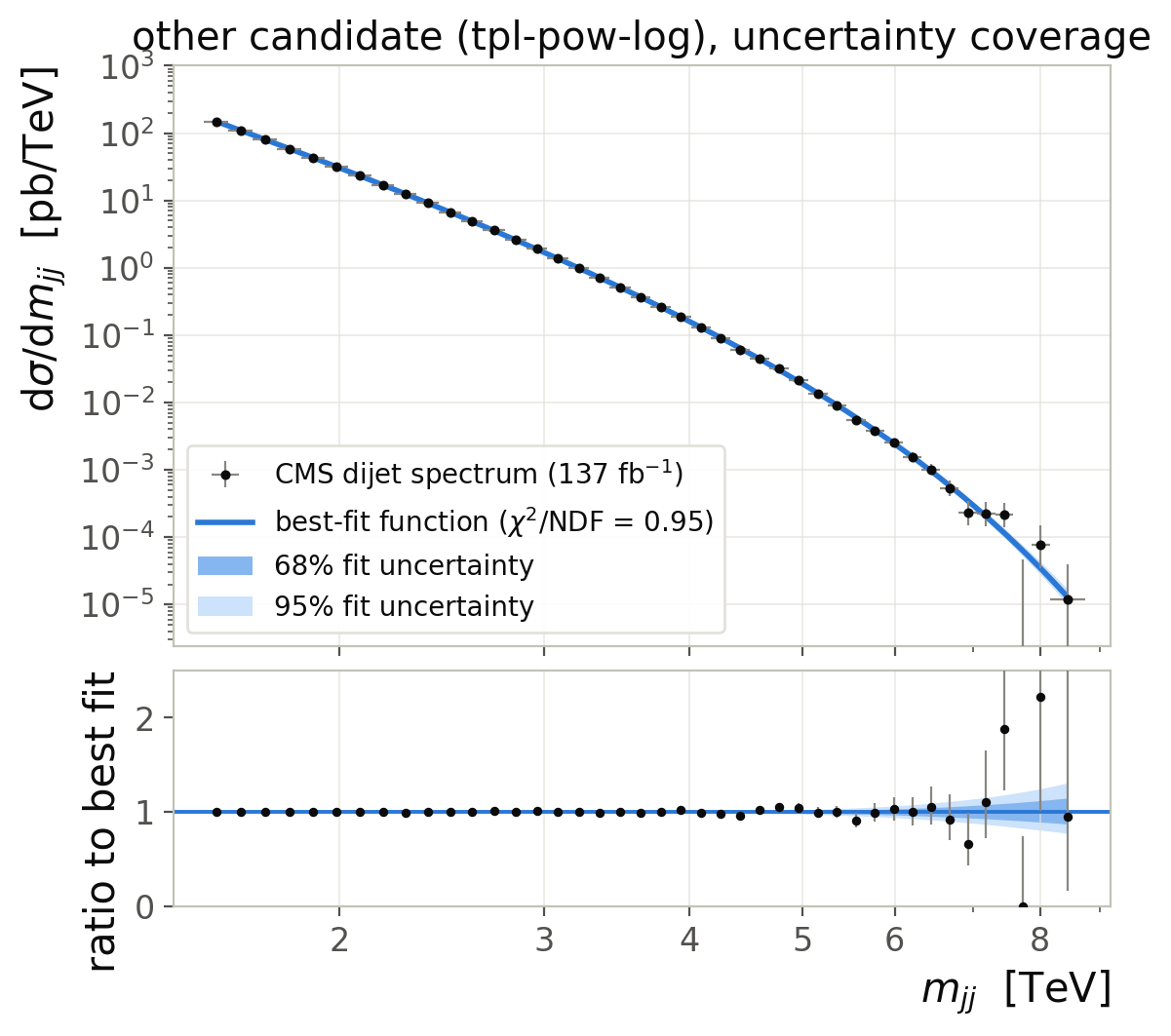}
  \includegraphics[width=0.38\textwidth]{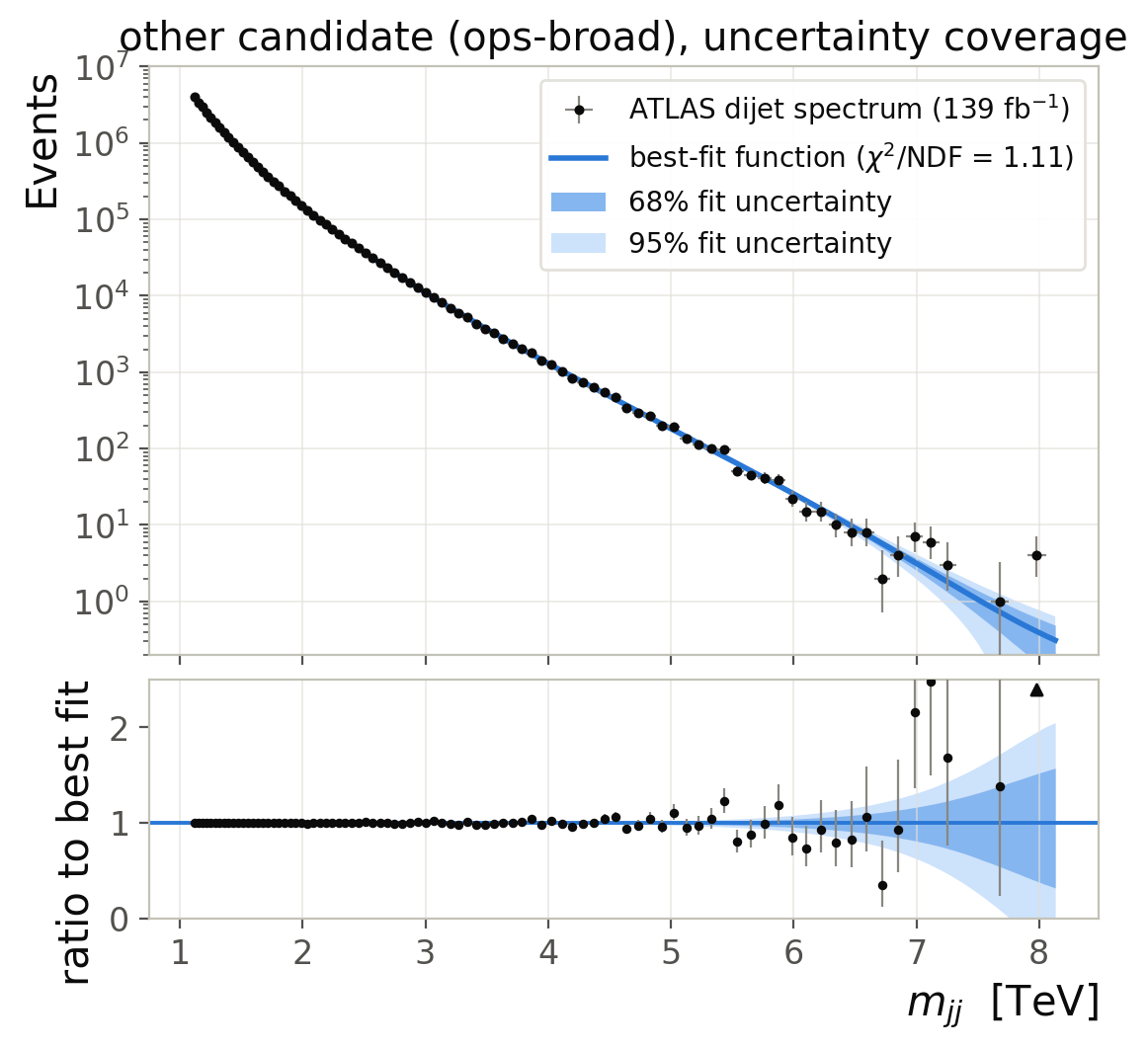}
  \caption{Two example candidate functions that are distinct from the dijet function and the UA2 function, compared to the CMS (left) and ATLAS (right) spectra, with 68\% and 95\% uncertainty bands propagated from the parameter covariance matrix. Lower panels show the ratio of data and bands to the best-fit function. The same uncertainty modeling is available for every candidate.}
  \label{fig:unc}
\end{figure}

Rediscovery is not itself a measure of usefulness, as the dijet function family was  derived empirically and is by no means the only form that can fit the spectra.
Instead, each run returned mostly non-dijet functions that model the spectra as well as the dijet function does.
Fig.~\ref{fig:chi2} shows the $\chi^2/\text{NDF}$ distribution of all candidate functions across the 560 runs.
Over 1000 candidates land at $\chi^2/\text{NDF}\approx 1$ from the 560 runs, and the rediscovered dijet and UA2 functions are only a tiny fraction of them.
Tab.~\ref{tab:candidates} lists the rediscovered candidates and one example candidate per search configuration that is distinct from the dijet and UA2 function families, all plotted in Fig.~\ref{fig:spectra}.
A wide variety of discovered forms, distinct from the dijet and UA2 functions, model the spectra equally well.
The majority of the rediscovered dijet forms have the $p_3\ln x$ term absent ($p_3=0$), and we have tested with a direct fit of the dijet function that returns $p_3$ consistent with zero for the CMS spectrum, while for the ATLAS spectrum the term improves the fit but not enough to justify its increased complexity in the search.
Each candidate returned by SymbolFit has uncertainty estimation embedded: Fig.~\ref{fig:unc} shows two non-dijet candidate functions plotted with their total uncertainty coverage, obtained via parameter sampling using the covariance matrix.

\begin{table}[t]
  \centering
  \scriptsize
  \setlength{\tabcolsep}{3.5pt}
  \renewcommand{\arraystretch}{1.1}
  \begin{tabular}{l l c | l c}
    \hline
     & \multicolumn{2}{c}{\textbf{CMS spectrum}} & \multicolumn{2}{|c}{\textbf{ATLAS spectrum}} \\
    \cline{2-3}\cline{4-5}
    Config & Candidate function & $\chi^2$/NDF & Candidate function & $\chi^2$/NDF \\ \hline
    rediscovered & $\dfrac{0.00672\,(1-x)^{8.15}}{x^{5.23}}$ & 0.93
                 & $\dfrac{0.802\,e^{-8.26\,x - 6.12\,x^{2}}}{x^{5.17}}$ & 1.43 \\
    \texttt{ops-broad} & $0.00255\left(\dfrac{\ln(0.72/x)}{x}\right)^{4.07}$ & 1.24
                       & $\left(-1.0 + \dfrac{0.826}{x + 0.00522}\right)^{5.62}$ & 1.13 \\
    \texttt{ops-log} & $\left(-\dfrac{0.176\ln x}{x}\right)^{x + x^{3.91} + \ln(1.4^{x}) + 4.27}$ & 0.92
                     & $\left(\dfrac{0.216^{x}}{x} - 0.315\right)^{5.11}$ & 1.22 \\
    \texttt{ops-log-tight} & $(4.05\,x)^{-7.35\,x - 6.11}$ & 1.18
                           & $\dfrac{\left(-1.32 + \dfrac{0.878}{x}\right)^{5.2}}{0.71 - x^{0.545/x}}$ & 1.23 \\
    \texttt{ops-minimal} & $\left(\dfrac{1.66}{x} - 2\right)^{5.61}\left(1.80\times 10^{-4} - \dfrac{4.11\times 10^{-6}}{x}\right)$ & 0.97
                         & $\dfrac{\left(\dfrac{0.757 - x}{x}\right)^{4.87}}{x + 0.0966}$ & 1.23 \\
    \texttt{tpl-exp-x} & $\exp\!\big[-(1.40 + \ln x)(1.03 + x)(5.95 + x)\big]$ & 1.15
                       & $\begin{array}{@{}l@{}} 3.07\exp\!\big[x - (2.95\,x + \ln x) \\ \qquad {}\times(3.94\,x^{2} - x + 4.83)\big] \end{array}$ & 1.13 \\
    \texttt{tpl-pow-log} & $(0.826 - 0.181\,x)^{32.2\ln x + 45.0}$ & 1.17
                         & $\big((x + 1.69)(2.59\,x^{4} + x)\big)^{-0.329\ln x - 7.04}$ & 1.16 \\
    \texttt{tpl-exp-log} & $\begin{array}{@{}l@{}} 1.01\exp\!\big[(2.63 - 0.0662\ln(1-x)) \\ \qquad {}\times(3\ln(1-x) - 2\ln x - 1.95)\big] \end{array}$ & 0.92
                         & $\dfrac{1.35\exp\!\big[(8.32 - \ln x)\ln(1-x)\big]}{x^{5.05}}$ & 1.22 \\
    \hline
  \end{tabular}
  \caption{The rediscovered functions (dijet for CMS, UA2 for ATLAS) and an example candidate function from each search configuration. Constants are best-fit values (uncertainties are not shown), rounded to three significant figures for readability ($\chi^2/\text{NDF}$ is computed from full precision). $x = m_{jj}/\sqrt{s}$. The same candidates are plotted in Fig.~\ref{fig:spectra}.}
  \label{tab:candidates}
\end{table}

\section{Summary}

Given only the observed dijet spectra published by CMS and ATLAS, SymbolFit rediscovers the very dijet and UA2 functions used in published searches, and returns many more functions that model the spectra equally well.
Our results suggest that the traditional parametric modeling method with trial-and-error selection of adequate functional forms and parameterizations, which is still the standard practice in HEP, can be replaced and automated by a machine in a data-driven way.
This can apply broadly to HEP data analyses wherever parametric modeling is performed.

\bibliographystyle{naturemag}
{\footnotesize
\bibliography{ref}
}

\end{document}